\documentclass[prl,twocolumn,amsmath,amssymb]{revtex4-1}
\usepackage{amsmath,amssymb,amsfonts,bm}
\usepackage{graphicx}
\usepackage{color}
\usepackage{bbold}
\usepackage{graphicx}
\usepackage{dcolumn}
\usepackage{epstopdf}
\usepackage{color}
\usepackage{epsfig}
\usepackage{url}
\usepackage{hyperref}
\usepackage{verbatim}
\usepackage{caption}
\usepackage{subcaption}

\newcommand{\be}{\begin{equation}}
\newcommand{\ee}{\end{equation}}
\newcommand{\ba}{\begin{aligned}}
\newcommand{\ea}{\end{aligned}}

\newcommand{\titleinfo}{Spectral statistics in spatially extended chaotic quantum many-body systems}

\graphicspath{ {./figures/} }

\begin{document}

\title{Spectral statistics in spatially extended chaotic quantum many-body systems}
\author{Amos Chan,  Andrea De Luca and J. T. Chalker}
\affiliation{Theoretical Physics, Oxford University, 1 Keble Road, Oxford OX1 3NP, United Kingdom}

\date{\today}

\begin{abstract}
We study spectral statistics in spatially extended chaotic quantum many-body systems, using simple lattice Floquet models without time-reversal symmetry. Computing the spectral form factor $K(t)$ analytically and numerically, we show that it follows random matrix theory (RMT) at times longer than a many-body Thouless time, $t_{\rm Th}$. We obtain a striking dependence of $t_{\rm Th}$ on the spatial dimension $d$ and size of the system. For $d>1$, $t_{\rm Th}$ is finite  in the thermodynamic limit and set by the inter-site coupling strength. By contrast, in one dimension $t_{\rm Th}$ diverges with system size, and for large systems there is a wide window in which spectral correlations are not of RMT form. Lastly, our Floquet model exhibits a many-body localization transition and we discuss the behavior of the spectral form factor in the localized phase.
\end{abstract}

\maketitle

A central theme in our understanding of highly excited states in complex quantum systems is that spectral properties are best discussed in statistical terms. First formulated in the framework of nuclear physics, this has led to the development of random matrix theory, in which one renounces knowledge of the Hamiltonian for a specific system and instead studies a statistical ensemble that is constrained only by symmetries \cite{Brody,Mehta}.

Random matrix theory (RMT) has proved to be of impressively wide applicability. It gives a precise description of many aspects of the quantum mechanics of low-dimensional chaotic systems \cite{Haake}, of mesoscopic disordered conductors \cite{Beenakker}, and of non-integrable many-body lattice models on the energy scale of the level spacing \cite{Montambaux}.

Spatially extended systems with local Hamiltonians offer a context in which it is natural that there should be limits on the applicability of RMT, since the combination of spatial structure and locality implies a preferred basis in Hilbert space, which is explicitly excluded from the theory. This is well understood for single-particle models of mesoscopic conductors in the diffusive regime. Here, characteristic scales are set by the Thouless time $t_{\rm Th}$, the time taken for a particle to explore the accessible phase space, and its energy counterpart $E_{\rm Th} \equiv \hbar/t_{\rm Th}$ \cite{Thouless}. With diffusion constant $D$ and linear size $L$, the time for a particle to cross such a system is $L^2/D$, so that $E_{\rm Th} = \hbar D/L^2$. An energy window of width $E_{\rm Th}$ contains many levels in a large system with dimension $d>2$, since the spacing between energy levels varies as $L^{-d}$. Spectral statistics follow RMT on scales much smaller than $E_{\rm Th}$, but have a different (although also universal) form on scales much larger than $E_{\rm Th}$ \cite{AltshulerShklovskii}. 

Our aim in this paper is to examine how spectral statistics are influenced by spatial dimension and system size in chaotic many-body systems with interactions that are local in space. We treat Floquet systems, both because they are convenient for our approach and because their lack of locally conserved densities is expected to result in particularly simple dynamics \cite{Huse2017,vonKeyserlingk2017a}. In the spirit of RMT, we study models drawn from an ensemble and compute their average properties. We treat systems without time-reversal symmetry. Important complementary results have been obtained recently for time-reversal symmetric systems within the framework of periodic orbit theory \cite{Prosen}, while previous numerical work \cite{Bertrand} on the ergodic phase in weakly disordered spin chains has identified a Thouless time growing with system size.

The focus of our discussion is the behaviour of the spectral form factor associated with the Floquet operator $W$, which generates time evolution over one cycle for a periodically driven quantum system. Denoting the eigenvalues of $W$ by $\{\theta_n\}$, the spectral form factor is defined for integer $t$ by
\begin{equation}
K(t) = \sum_{m,n} \langle e^{it(\theta_m - \theta_n)} \rangle\,,
\end{equation}
where $\langle \ldots \rangle$ indicates the ensemble average. For a system without spatial structure or time-reversal symmetry, $W$ is modelled within RMT by a unitary $N\times N$ matrix drawn from the Haar distribution. In this case \cite{Mehta}
\begin{equation}
\label{RMTform}
K(t) = N^2 \delta_{t,0}+\left\{ \begin{array}{ll}
 |t| &\quad 0< |t| \leq N\\ 
N &\quad N\leq |t| \end{array}\right.
\end{equation}
The linear variation of $K(t)$ with $t$ for $0<t\leq N$ is a consequence of spectral rigidity, and the saturation time $t=N$ is set by the inverse level spacing.

\begin{figure}[tb]
	\includegraphics[width=0.5\textwidth]{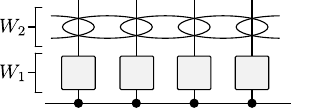}
	\caption{Diagrammatic representation of the Floquet operator, where $W_1$ contains unitary 1-gates, and $W_2$ contains diagonal 2-gates with  random phases.}
	\label{fig:0}
\end{figure}

We define the Thouless time to be the time-scale beyond which $K(t)$ has RMT form. Consider a many-body system consisting of $L^d$ chaotic grains with Hilbert space dimension $q$ for each grain. If the grains are uncoupled one has $K(t) = t^{L^d}$ for $0<t \leq q$. If the grains are weakly coupled it is natural to expect that the coupling is effective only at long times, so that $K(t) \gg t$ for $0< t \ll t_{\rm Th}$ and $K(t) =t$ for $t_{\rm Th} \ll t \leq q^{L^d}$. In the following we define a simple model of this kind, show that it exhibits just this behaviour, and determine the dependence of $t_{\rm Th}$ on dimension and system size. We use analytical calculations for large $q$ and general $d$, and  numerical simulations for small $q$ and $d=1$. We also present qualitative arguments in support of the idea that our main results are generic.

The model we study consists of $q$-state `spins' arranged with nearest-neighbour coupling on a $d$-dimensional lattice. To be specific, we describe it first for $d=1$ and $L$ sites. We use site labels $n=1 \ldots L$ and orbital labels $a_n = 1 \ldots q$ on the $n$th site. The $q^L\times q^L$ Floquet operator $W=W_2\cdot W_1$ is a product of two factors. $W_1 = U_1 \otimes U_2 \otimes  \ldots U_L$ generates independent rotations at each site $n$, with $q\times q$ unitary matrices $U_n$ chosen randomly and independently from the Haar distribution. $W_2$ couples neighbouring sites and is diagonal in the basis of site orbitals  (Fig.~\ref{fig:0}). The phase of the diagonal elements is a sum of terms depending on the quantum states of adjacent sites, so that
\begin{equation}\label{coupling}
[W_2]_{a_1, \ldots a_L;a_1, \ldots a_L} = 
\exp\left(i\sum_n \varphi_{a_n,a_{n+1}}\right)\,.
\end{equation}
We take each $\varphi_{a_n,a_{n+1}}$ to be an independent Gaussian random variable with 
mean zero and variance $\epsilon$.
To construct a similar model in higher dimensions we use sites $\bf r$ on a hypercubic lattice and take independent contributions $\varphi_{a_{\bf r},a_{\bf r}^\prime}$ to the phase of $W_2$ from each neighbouring pair of sites, $\bf r$ and ${\bf r}^\prime$. 

The model is distinguished from a related model that we have investigated elsewhere \cite{previous} in having two separate parameters: $q$ and $\varepsilon$. This brings the advantage for analytical work that one can use large $q$ to control calculations, but retain variable coupling between sites \footnote{Behaviour at times $t< t_{\rm Th}$ in the model of Ref.~\cite{previous} is hard to access as it is sub-leading in a $1/q$ expansion.}. Sites are independent for $\varepsilon=0$ and are maximally coupled for $\varepsilon \gg 1$. It is natural to expect two phases: a many-body localised (MBL) phase  \cite{BAA,MBLreview1,MBLreview3,MBLreview2} for $\varepsilon < \varepsilon_c(q)$ and an ergodic phase \cite{LuitzReview} for $\varepsilon > \varepsilon_c(q)$. We are concerned here with behaviour in the ergodic phase; we believe (see below) that in one dimension: (i) $\varepsilon_c(q) \to 0$ as $q\to \infty$, (ii) the model is MBL for all $\varepsilon$ at $q=2$, and (iii)  $\varepsilon_c(3) \approx 0.25$. In numerical studies of the ergodic phase we use $q=3$ and $\varepsilon =1$.

The spectral form factor is related to the Floquet operator via
\begin{equation}
K(t) = \langle [{\rm Tr}\, W(t)][{\rm Tr}\, W^\dagger(t)]\rangle\,,
\end{equation}
where we use $W(t)$ to denote the $t$-th power of $W$.

It is possible to evaluate $K(t)$ exactly for fixed $t$ and $\varepsilon$ in the limit $q\to\infty$, via a mapping to a Potts model that we describe shortly. In one dimension (with periodic boundary conditions) we find
\begin{equation}\label{K1d}
K(t) = (t-1)(1 - e^{-\varepsilon t})^L + [1+(t-1)e^{-\varepsilon t}]^L\,.
\end{equation}
This form can be simplified in the regime of most interest, $t\gg 1$ and $L \gg 1$, giving
\begin{equation}\label{Kasymptotic}
K(t) \sim \left\{\begin{array}{lll} t^{L/\xi(t)}& \quad & t\ll t_{\rm Th}\\
t & & t\gg t_{\rm Th}\,,
\end{array}\right.
\end{equation}
with
\begin{equation}\label{xi}
\xi(t) = 
\frac{\ln t}{\ln[1 + (t-1)e^{-\varepsilon t}]} \sim \frac{\ln t}{t}e^{\varepsilon t}
\end{equation}
(where the asymptotic form is for $e^{\varepsilon t} \gg t$) and
\begin{equation}
t_{\rm Th} = \frac{\ln L}{\varepsilon}\,.
\end{equation}
The behaviour of $K(t)$ at times $t\ll t_{\rm Th}$ has a clear interpretation: the spectral fluctuations are the same as if the system consisted of $L/\xi(t)$ uncoupled pieces. The size $\xi(t)$ of each piece diverges very rapidly with $t$, and the Thouless time $t_{\rm Th}$ is the time at which $\xi(t)$ reaches the system size $L$.

In higher dimensions $d > 1$, we obtain for $t\gg1$ and $\varepsilon \ll 1$ (but $\varepsilon t$ arbitrary)
\begin{equation}\label{Kd>1}
K(t) = \left\{\begin{array}{ccc} (te^{-d\varepsilon t})^{L^d} & \quad & t\leq t_{\rm Th}\\
t & & t\geq t_{\rm Th}
\end{array}\right.
\end{equation}
where the Thouless time is the larger of the two solutions to $t_{\rm Th} e^{-d \varepsilon t_{\rm Th}} = 1$. Hence for $d>1$ there is a fixed value for $t_{\rm Th}$, which is independent of system size and is large if the inter-site coupling $\varepsilon$ is small.

Eqns.~(\ref{Kasymptotic}--\ref{Kd>1}) constitute the main analytical results of this paper. Before outlining their derivation, we present computational studies probing behaviour at small $q$. Our numerical methods are outlined and further results are given in {\it supplemental material}~\footnote{See supplementary material at [url].}.

\begin{figure}[tb]
\includegraphics[width=0.5\textwidth]{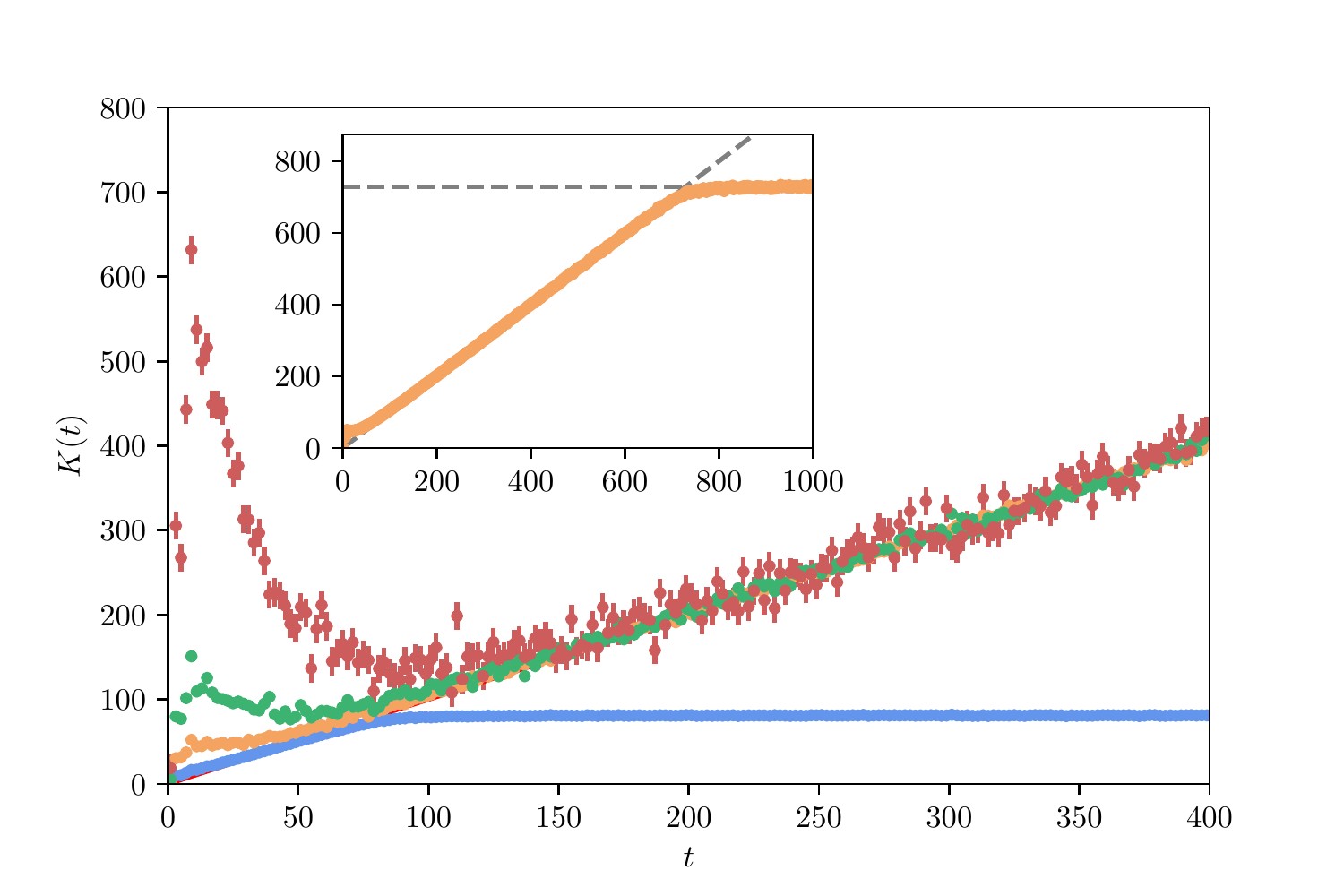}
\caption{$K(t)$ at small $t$, showing deviations from RMT form that grow with $L$. Data for $L=4$, $6$, $8$ and $10$ (from bottom to top). Inset: $K(t)$ vs $t$ for $L=6$.}
\label{fig:2}
\end{figure}

\begin{figure}[tb]
\includegraphics[width=0.5\textwidth]{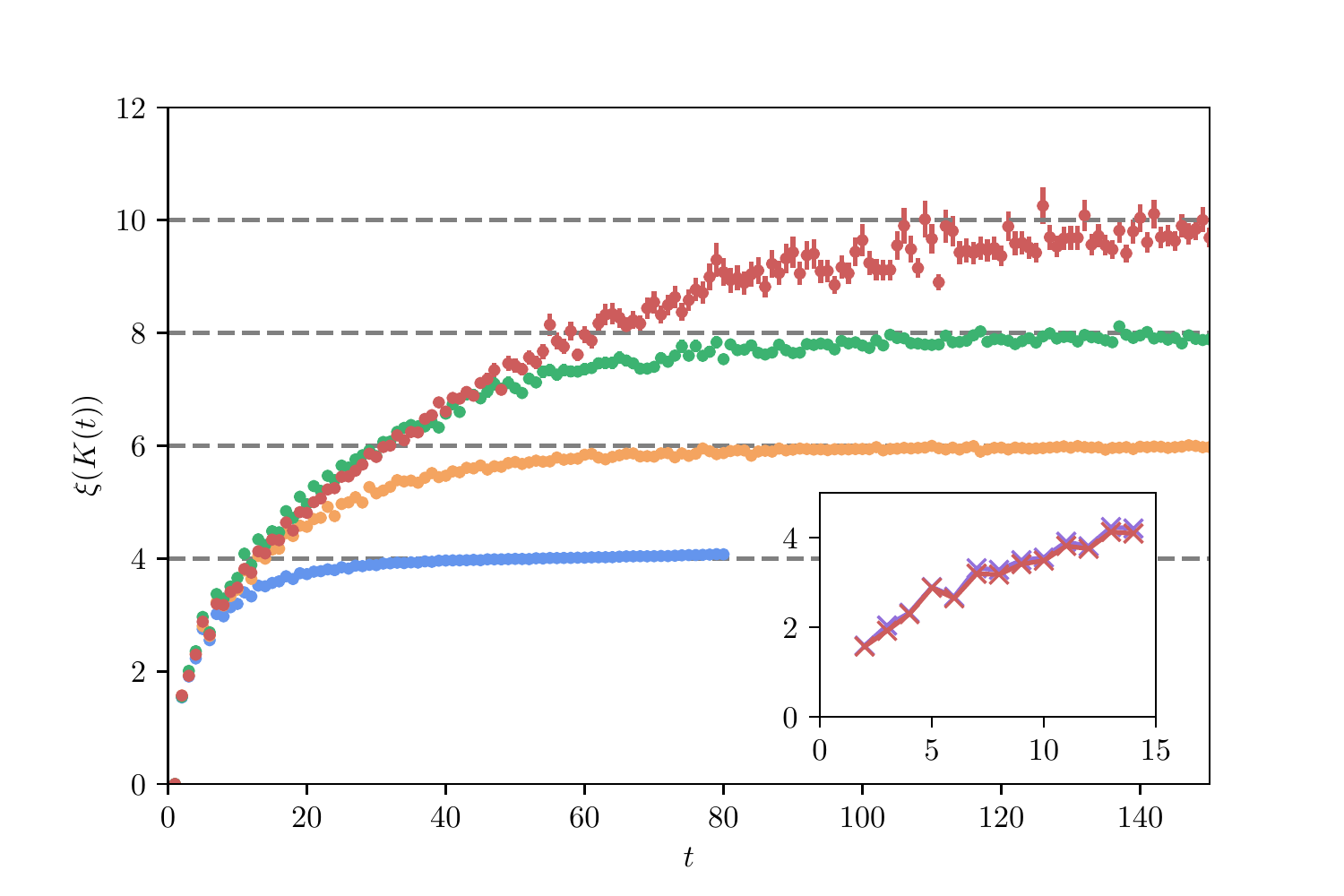}
\caption{$\xi_L(t)$ vs $t$. Main panel: $L=4$, $6$, $8$ and $10$ (from bottom to top). Inset: $L=10$ (red) and $L=100$ (purple) at short times.}
\label{fig:3}
\end{figure}

The behaviour of $K(t)$ is shown in Fig.~\ref{fig:2} for $q=3$, $\varepsilon=1$ \footnote{$K(t)$ shows small period $\Delta t = 2$ oscillations. In Fig.~\ref{fig:2} we display data only for odd $t$. Even-$t$ data show similar results. Both sets are included in Fig.~\ref{fig:3}.}. The inset gives an overview of how $K(t)$ approximates the RMT form given in Eq.~\eqref{RMTform}. Departures from this are largest at early times and grow with increasing system size, as illustrated in the main panel of Fig.~\ref{fig:2}. To interpret these deviations, we adapt Eq.~(\ref{xi}) to finite system size and express spectral fluctuations in terms of the length scale $\xi_L(t) = (L\ln t)/\ln K(t)$. As we demonstrate in Fig.~\ref{fig:3}, $\xi_L(t)$ appears to be independent of $L$ at times sufficiently short that $\xi_L(t) \ll L$. It grows rapidly with increasing $t$ and saturates at $\xi_L(t) = L$ for $t$ large (but less than $q^L$). The timescale for saturation is the Thouless time; it increases rapidly with $L$ over the available range of system sizes. We believe these numerical results provide a good indication that qualitative behaviour throughout the ergodic phase in $d=1$ is similar to what we have obtained analytically at large $q$, although (as we discuss below) the specific functional form of $\xi_L(t)$ in Fig.~\ref{fig:3} appears not to be very close to that in Eq.~(\ref{xi}).

\begin{figure}[tb]
	\includegraphics[width=0.45\textwidth]{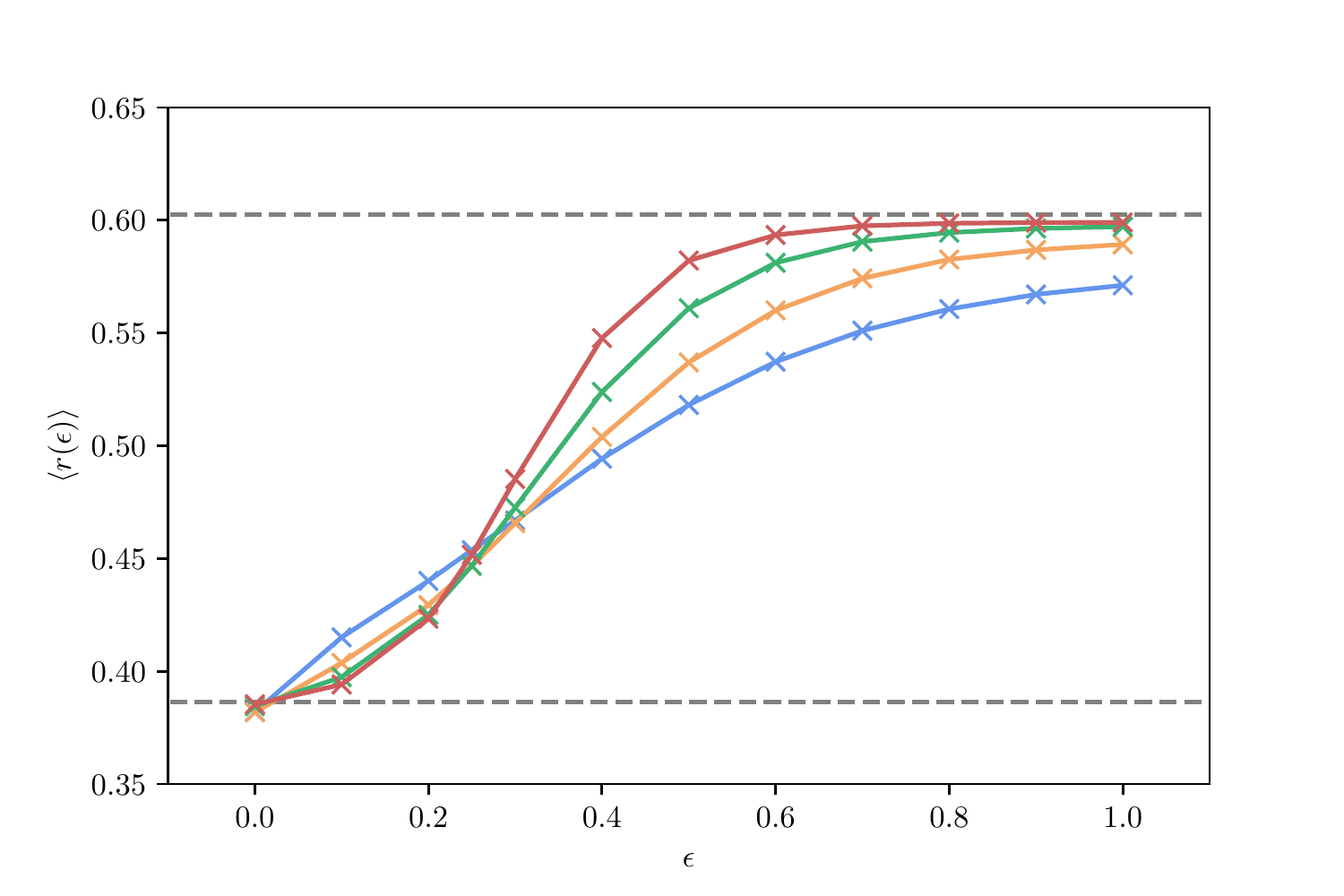}
	\caption{Identification of MBL and ergodic phases: $\langle r \rangle$ vs $\varepsilon$ for $q=3$ with system sizes $L=4, 5, 6, 7$. Upper and lower horizontal lines indicate the values expected in an ergodic \cite{Atas} and an MBL phase \cite{OganesyanHuse}, respectively.}
	\label{fig:4}
\end{figure}

We close with a discussion of the phase diagram of the model as a function of $\varepsilon$ for $q=2$ and $3$.
We use the well-established diagnostic computed from the ratio of eigenphase spacings: let $\theta_{n-1}$, $\theta_n$ and $\theta_{n+1}$ be three successive eigenphases, and define $r={\rm min}\{\theta_n- \theta_{n-1},\theta_{n+1}-\theta_n\}/{\rm max}\{\theta_n- \theta_{n-1},\theta_{n+1}-\theta_n\}$ \cite{OganesyanHuse}. Then $\langle r \rangle$ takes different characteristic values in the two phases. 
	In Fig.~\ref{fig:4},
we provide clear numerical evidence for a critical point at $\varepsilon_c(q=3) \approx 0.25$. For $q=2$, behaviour of $K(t)$ and $\langle r \rangle$  indicate that there is only an MBL phase \cite{Note2}. 

The behaviour of $K(t)$ is quite different in the MBL phase. Whereas in the ergodic phase $K(t)$ reaches its limiting value of $q^L$ at a time $t=q^L$ set by the inverse level spacing, in the MBL phase $K(t)$ approaches $q^L$ at a time that remains finite as $L\to \infty$ \cite{Note2}.

We now turn to the derivation of our analytic results, Eqns.~(\ref{Kasymptotic}--\ref{Kd>1}), which are based on an evaluation of $\langle [{\rm Tr}\, W(t)][{\rm Tr}\, W^\dagger(t)]\rangle$ at large $q$. 

Consider first the average on $W_1$, and in particular, the Haar unitary $U_n$ at site $n$. The factors involved have the form
\begin{equation}\label{Us}
\begin{array}{l}
\langle [U_n]_{a(1),a(2)} [U_n]_{a(2),a(3)} \ldots [U_n]_{a(t),a(1)}\times \qquad \\  \qquad \times [U_n]^*_{b(1),b(2)} 
[U_n]^*_{b(2),b(3)} 
\ldots [U_n]^*_{b(t),b(1)}\rangle\,.  
\end{array}
\end{equation}
The leading contributions to this average are the ones in which the indices $b(1),b(2) \ldots b(t)$ are a cyclic permutation of the indices $a(1),\,a(2) \ldots a(t)$ \cite{Samuel,Mello,Brouwer1996}. Their average at leading order is $q^{-t}$; corrections and other contributions are smaller by inverse powers of $q$. In the large-$q$ limit we hence need to consider only these cyclic permutations, which can be specified by the relative shift $s$, with $a(r) = b(r+s)$ for $r=1, 2 \dots t$, where $r$ and $r+s$ are counted mod $t$. Summing over values for $\{a(r)\}$ at fixed $s$ generates a factor of $q^t$, cancelling the factor of $q^{-t}$ from the average. The shift takes the values $s=1, 2 \ldots t$ and serves to label at each site the dominant contributions that remain after averaging on $W_1$. 

Next consider the average on $W_2$, and specifically the contributions to the phase in (\ref{coupling}) arising from a particular pair of neighbouring sites, $n$ and $n^\prime$, for the leading terms that are selected via the average on $W_1$. Denoting indices on $U_n$ by $a(r)$ and those on $U_{n^\prime}$ by $a^\prime(r)$, and the shifts by $s$ and $s^\prime$, this phase contribution is
\begin{equation}
\Phi_{n,n^\prime} =\sum_{r=1}^{t} (\varphi_{a(r),a^\prime(r)} - \varphi_{a(r+s),a^\prime(r+s^\prime)})\,.
\end{equation}
For $s=s^\prime$, terms cancel in pairs and $\Phi_{n,n^\prime}$ vanishes. Conversely, for $s\not= s^\prime$ at large $q$ and fixed $t$, all except a fraction ${\cal O}(q^{-1})$ of the possible assignments for $\{a(r)\}$ and $\{a^\prime(r)\}$ leave $2t$ independent terms in $\Phi_{n,n^\prime}$. As a result, $\langle e^{i \Phi_{n,n^\prime}} \rangle$ has the value $1$ if $s=s^\prime$, and $e^{-\varepsilon t}$ if $s\not= s^\prime$. 

Combining these results, we see that $K(t)$ at large $q$ is given \emph{exactly} by the partition function of a classical $t$-state ferromagnetic Potts model, on the same lattice as in the original many-body quantum system. This partition function is a sum over configurations in which the shifts at each site independently take one of $t$ possible states, with factors in the `Boltzmann weight' from each coupled pair of sites that are $1$ or $e^{-\varepsilon t}$, according to whether the sites of the pair are in the same or different states.

In $d=1$ it is straightforward to evaluate the partition function using a transfer matrix, yielding Eq.~(\ref{K1d}). For $d>1$, we note that the main interest is in behaviour at $t\gg1$ and that the $t$-state ferromagnetic Potts model in $d>1$ has an ordering transition that is strongly discontinuous at large $t$ \cite{Baxter}. In the disordered phase ($e^{-\varepsilon t}$ close to 1) the state of each site fluctuates almost independently of its neighbours, while in the ordered phase nearly all sites occupy the same state. The two phases give the two contributions to Eq.~(\ref{Kd>1}).

An important question is whether this large-$q$ behaviour describes more broadly ergodic many-body Floquet systems without time-reversal symmetry. 
We believe that our solution in fact suggests an appealing and potentially general picture, as follows. Note first that an expansion of $[{\rm Tr}\, W(t)][{\rm Tr}\, W^\dagger(t)]$ can be expressed for general $W$ as a sum over terms arising from the product of two $t$-step orbits in the space of many-body states, one from expansion of ${\rm Tr}\, W(t)$ and the other from ${\rm Tr}\, W^\dagger(t)$. Contributions in which one orbit is a cyclic permutation of the other are always real and positive, while other contributions may have any phase. It is therefore natural to anticipate that orbits paired in this way will dominate under many circumstances. Just as with periodic orbit theory for the quantisation of low-dimensional chaotic systems \cite{Berry}, or diffusons in mesoscopic conductors \cite{Khmelnitskii}, we expect these paired orbits to give the RMT form for $K(t)$ at $t<N$. 

In a spatially extended system, a clear alternative to a global pairing of orbits is that orbits are locally paired in this way, but with a cyclic permutation that varies between regions separated by domain walls. The simplifying features of our model at large $q$ make this picture exact and restrict domain walls to run only in the time direction. A more general treatment should include local disruptions to the pairing, and domain walls with locations that fluctuate in both space and time. Equivalence to a $t$-state Potts model may survive if these additional fluctuations are not qualitatively important, and provided the statistical cost of domain walls remains proportional to their $d-1$ dimensional area. 

Domain walls between pairings, similar to the ones we discuss here, have been proposed previously in discussions of entanglement spreading in random unitary circuits \cite{Nahum2017,Nahum2017a,vonKeyserlingk2017}, and an equivalent construction emerged in our study of entanglement in Floquet systems \cite{previous}.

While the spirit of our approach and aspects of our results have much in common with a recent evaluation \cite{Prosen} of $K(t)$ for a Floquet spin chain using periodic orbit theory, there are also important differences. In particular, the significance of locality  and dimensionality, and the notion of domain walls separating different pairings of cycles seem specific to the present work. The logarithmic scaling of the Thouless time obtained in \cite{Prosen} appear via a different mechanism, which in outline is as follows. Periodic orbit theory yields $K(t) = (2t/\beta) \sum_{a} P_{aa}(t)$ for $0<t \ll N$, where $\beta= 1$ or $2$ in systems with or without time-reversal symmetry. Here $P_{ab}(t)$ is the probability within the diagonal approximation for propagation in $t$ time steps between states $a$ and $b$ from the many-body Hilbert space. It is given by $P_{ab}(n) =  [T^n]_{ab}$ with $T_{ab} = |W_{ab}|^2$. Let the eigenvalues of $T$ be $\lambda_1, \lambda_2 \ldots$, with $|\lambda_1| \geq |\lambda_2| \geq \ldots$. Unitarity of $W$ ensures that $\lambda_1=1$. RMT behaviour follows if $\lambda_2 = 0$, and the corrections of \cite{Prosen} arise because the subleading eigenvalues of $T$ are not zero. In our model $\lambda_2 \sim {\cal O}(q^{-1})$ and so corrections of this type are absent for $q\to\infty$. 
We note that corrections to RMT similar to those of Ref.~\cite{Prosen} have recently been found in the SYK and other models \cite{ShenkerGroup2018}; we think it is likely that they arise by the same mechanism as in \cite{Prosen}, which (in contrast to ours at large $q$) is dimension-independent. 

By contrast, many-body systems with a locally conserved density that relaxes diffusively with diffusion constant $D$ are expected \cite{ShenkerGroup2018,future2} to have $t_{\rm Th} \sim L^2/D$, as in the singe-particle case. We expect ergodic time-independent Hamiltonians to belong to this category, in consequence of their conserved energy density.

The exponential divergence of $\xi(t)$ with $t$ at large $q$ is apparently more rapid than the growth shown for $q=3$ in Fig.~\ref{fig:3}. We expect at finite $q$ that rare weak links in a disordered one-dimensional system \cite{Nahum2017b} change the growth of $\xi(t)$ at long times from exponential to power-law. A weak link arises in our model between sites $n$ and $n+1$ if the coupling phase $\varphi_{a_n,a_{n+1}}$ has approximately the separable form: $\varphi_{a_n,a_{n+1}}= \tilde{\varphi}_{a_n} + \tilde{\varphi}_{a_{n+1}} + {\cal O}(\delta)$. This condition imposes $(q-1)^2$ constraints on the entries of $\varphi_{a_n,a_{n+1}}$. It is therefore satisfied in the ensemble to an accuracy $\delta\ll 1$ with probability $\propto \delta^{(q-1)^2}$. We expect the value of $\delta^2$ will play a similar role to that of $\varepsilon$ in Eq.~(\ref{K1d}), so that links of strength $\delta$ are ineffective until a time $t \sim \delta^{-2}$, implying $\xi(t) \lesssim t^{(q-1)^2/2}$ for large $t$. This upper bound is compatible with our analytical results, since the exponent diverges for $q\to\infty$. The bound appears to rise faster with $t$ than our data at $q=3$, indicating either large finite-size effects or the existence of another, more effective, mechanism for the formation of weak links.


A number of important points remain open. First, it is unclear how or whether our definition of a Thouless energy in terms of the spectral form factor is related to alternative definitions involving properties of matrix elements \cite{Luitz2016,Serbyn2016,Serbyn2017,LuitzReview}. More broadly, it will be valuable if the description of many-body quantum dynamics in terms of paired orbits can be developed further.


We thank Adam Nahum for extensive discussions. The work was supported in part by EPSRC Grant No. EP/N01930X/1.

\bibliography{FloquetChaos}

\onecolumngrid
\newpage 

\appendix
\setcounter{equation}{0}
\setcounter{figure}{0}
\renewcommand{\thetable}{S\arabic{table}}
\renewcommand{\theequation}{S\arabic{equation}}
\renewcommand{\thefigure}{S\arabic{figure}}

\begin{center}
	{\Large Supplementary Material \\ 
		\titleinfo
	}
\end{center}
In this supplementary material we provide additional details about:
\begin{itemize}
	\item Numerical methods employed in the calculation of $K(t)$ and $\xi(t)$.
	\item Numerical results for $K(t)$ and $\xi(t)$ in a Floquet model that has an ergodic phase at $q=2$. This model was studied analytically at large $q$ in \cite{previous}. At $q=2$ it gives access to the ergodic phase for larger system sizes than the principal model we consider, which has only an MBL phase at $q=2$.
		\item Numerical results for $K(t)$ and $\langle r \rangle$ for the many-body localized phase at $q=2$. 
\end{itemize}

\section{Numerical methods}
Our numerical results for $q=3$ are obtained with three complementary methods. We use (i) exact diagonalisation (ED) of $W$ to obtain $K(t)$ at all $t$ for $L\leq 6$. Since it is faster to act with the quantum circuit on a vector than it is to do ED, there are alternatives that can treat larger systems. We use: (ii) Monte Carlo (MC) to compute ${\rm Tr}\, W(t)$ by evaluating $\langle u | W(t) | u \rangle$ for randomly chosen vectors $|u\rangle$, giving access to $L\leq 10$ at times of most interest ($|t|\leq 500$); and (iii) a transfer matrix (TM) acting in the space direction, giving access to very large systems ($L \leq 100$) but only for short times ($|t| \leq 14$). With each method we average over a large number ${\cal N}$ of realisations: ${\cal N} \sim 10^5$ for ED; ${\cal N} \sim 10^6$ for MC; ${\cal N} \sim 10^3$ for TM. 

Our numerical results for $q=2$ are obtained using ED with ${\cal N} \sim 10^5$.
{
	\section{Spectral form factor for an alternative Floquet model for chaotic quantum many-body systems}
	Here we present  numerical results for $K(t) $ and $\xi(t)$ in a Floquet model that was studied analytically at large $q$ in \cite{previous}. This model is defined as a Floquet random unitary quantum circuit acting on $q$-state `spins'  on a 1-dimensional lattice with an even number of sites $L$. The $q^L\times q^L$ Floquet operator $W=W_2\cdot W_1$ is a product of two factors, $W_1 = U_{1,2} \otimes U_{3,4} \otimes  \ldots U_{L-1,L}$ and $W_2 = \mathbb{1}_q \otimes U_{2,3} \otimes U_{4,5} \otimes  \ldots\mathbb{1}_q $. Each $U_{i,i+1}$ is a $q^2 \times q^2$ unitary matrix acting on the Hilbert space of the spins at sites $i$ and $i+1$. The matrices $U_{i,i+1}$ are chosen randomly and independently from the Haar distribution, and $\mathbb{1}_q$ denotes $q \times q$ unit matrix. 
	
	We use ED to calculate $K(t)$ and $\xi_L(t) = \log (t) / (L \log K(t))$ for $q=2$ and $4\leq  L \leq 14$, taking $\mathcal{N} \sim 10^3$. The data for $K(t)$ (Fig.~\ref{FRUQC_K}) show a clear deviation from the RMT behaviour for early time. The behaviour of $\xi(t)$ in this model (Fig.~\ref{FRUQC_Xi}) is similar to that for the model discussed in the main text. 
	
	\begin{figure}[h!]
		\centering
		\includegraphics[angle=0, width=0.5\textwidth]{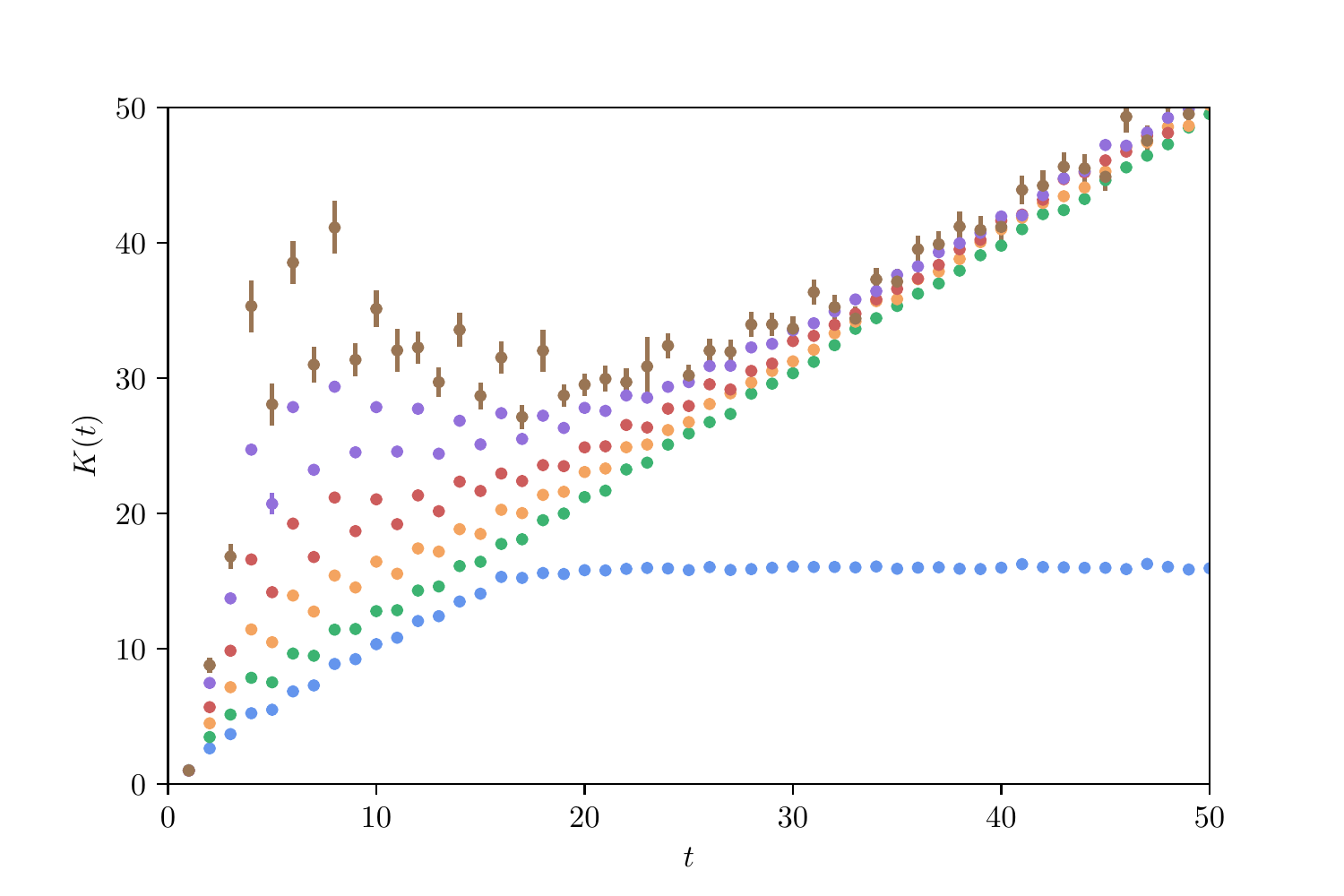}
		\caption{$K(t)$ at small $t$, showing deviations from RMT form that grow with $L$. Data for $L=4,6, \dots, 14$ (from bottom to top).}\label{FRUQC_K}
	\end{figure}\label{supp1}
	\begin{figure}[h!]
		\centering
		\includegraphics[angle=0, width=0.5\textwidth]{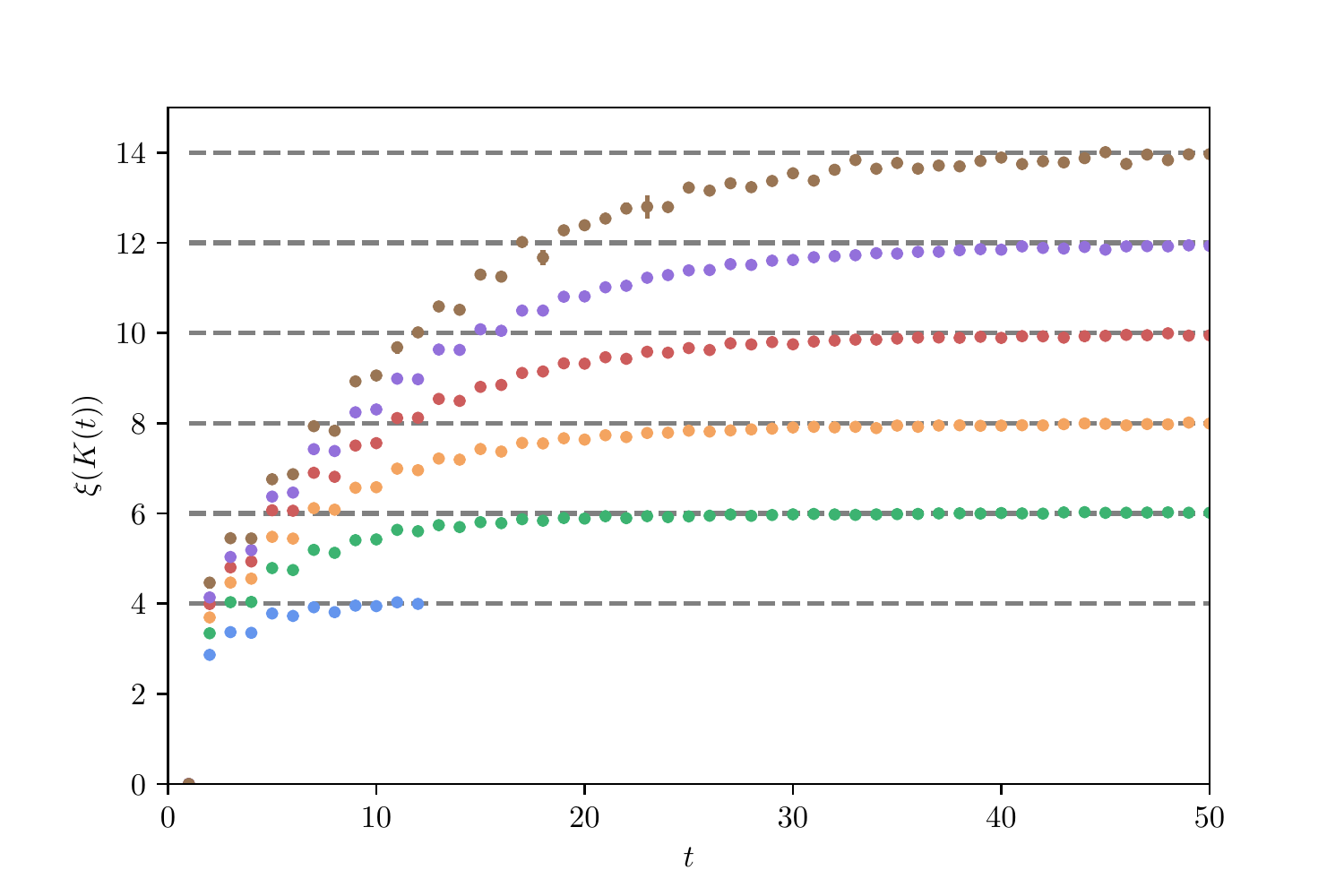}
		\caption{$\xi_L(t)$ vs $t$ for $L=4,6, \dots, 14$ (from bottom to top). }  \label{FRUQC_Xi}
	\end{figure}


}

{
	\section{Numerical results for $K(t)$ and $\langle r \rangle$ for the many-body localized phase at $q=2$}
	Using ED with $\mathcal{N} \sim 10^5$ realizations, we compute $K(t)$ and $\langle r \rangle$ in Fig.~\ref{fig:5}. Both $K(t)$ and $\langle r \rangle$ show a  clear deviation from expected RMT forms and suggest that there is only an MBL phase (see the main text for discussion).

	\begin{figure}[h!] 
		\centering 
		\begin{subfigure}{.5\textwidth}
			\centering
			\includegraphics[angle=0, width=1\linewidth]{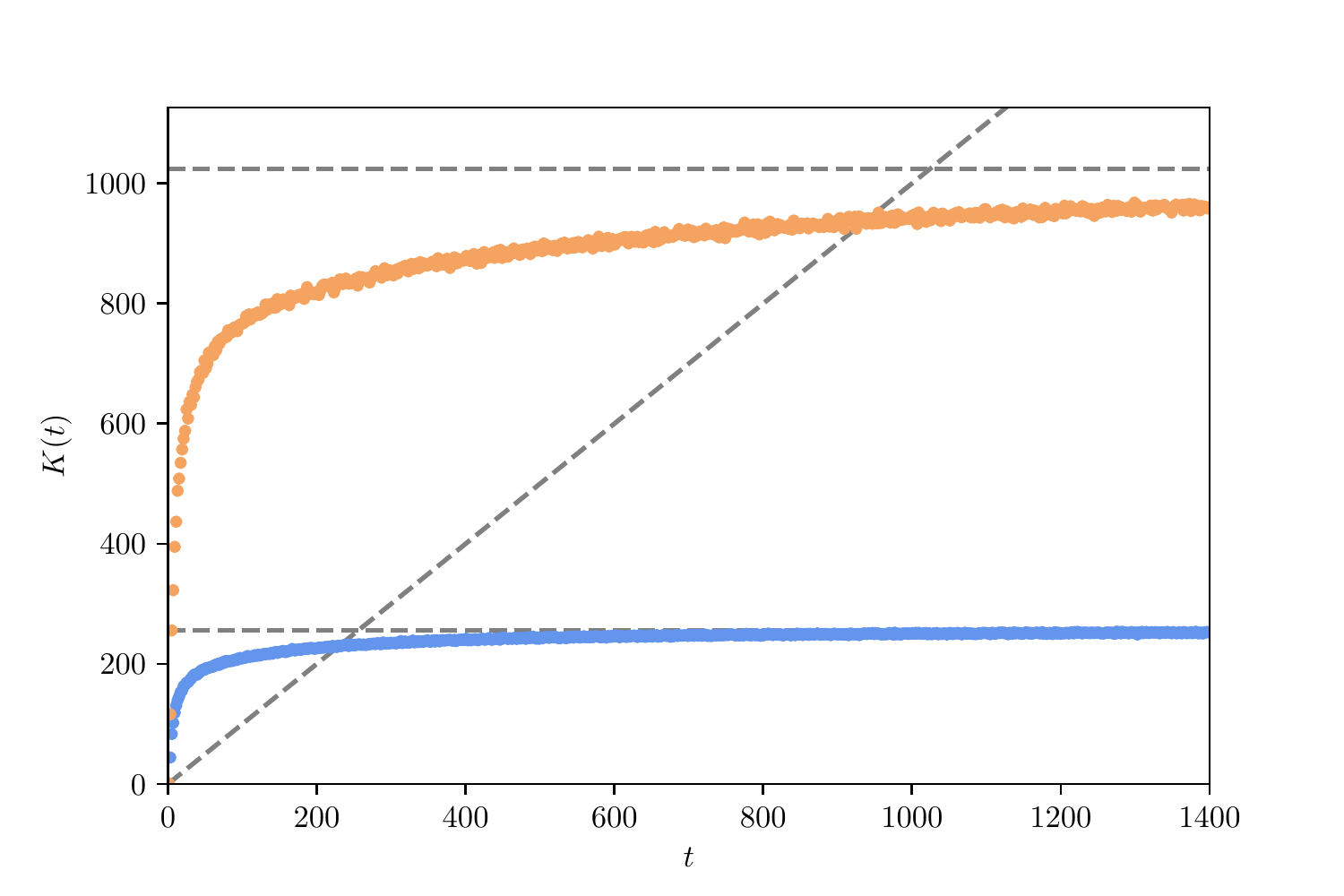}
			\caption{ }
		\end{subfigure}%
		\begin{subfigure}{.5\textwidth}
			\centering
			\includegraphics[angle=0, width=1\linewidth]{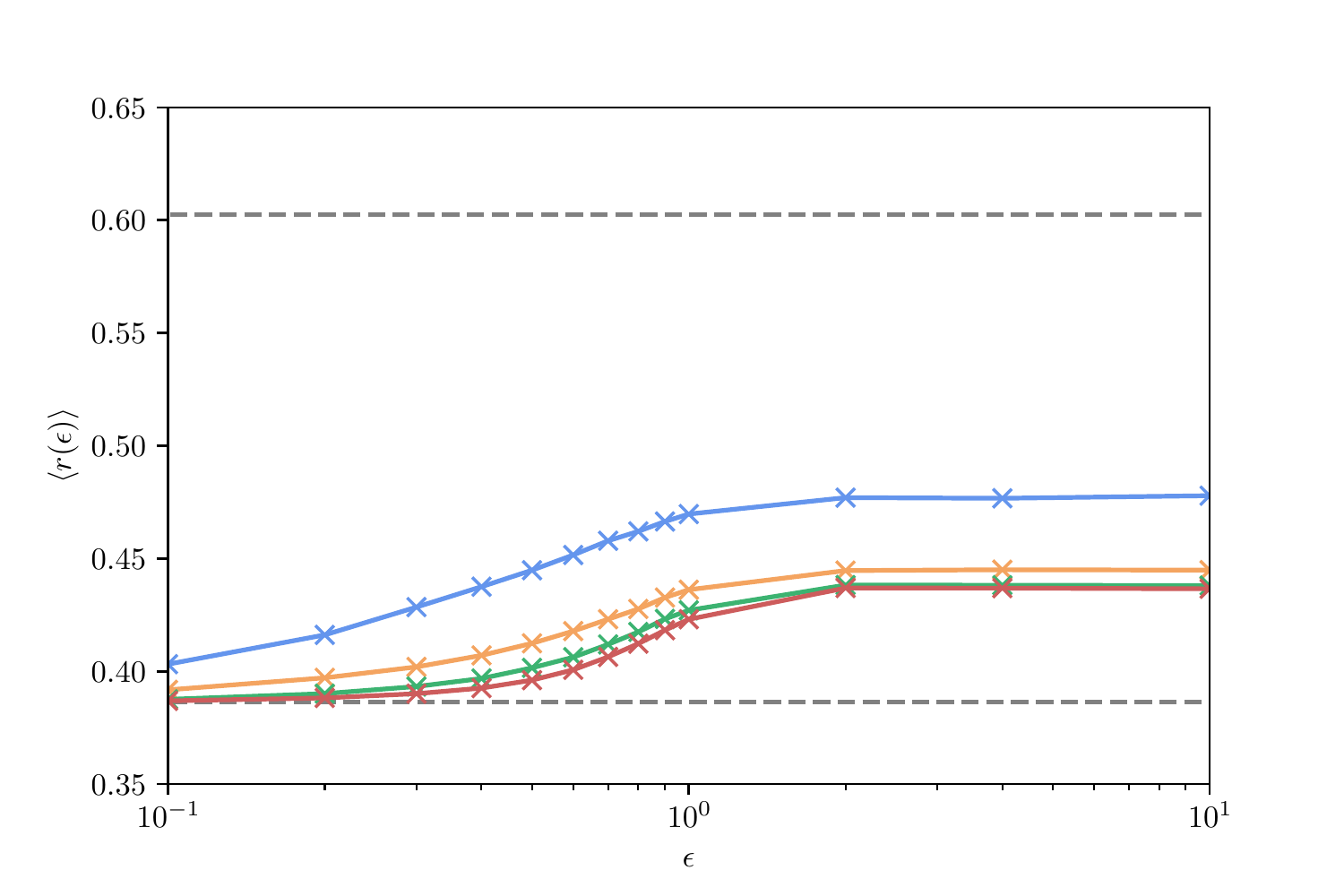}
			\caption{ }
		\end{subfigure}
		\caption{On the left, $K(t)$ vs $t$ in the MBL phase for $q=2$ and $\varepsilon=1$ for $L=8$ and $10$. On the right, $\langle r \rangle$ vs $\varepsilon$ for $q=2$; $L=4,6,8$ and 10, from top to bottom.} \label{fig:5}
	\end{figure}
}

\end{document}